\documentclass{ws-procs9x6}
\newcommand{\begit}{\begin{itemize}}
\newcommand{\enit}{\end{itemize}}
\newcommand{\begen}{\begin{enumerate}}
\newcommand{\enen}{\end{enumerate}}
\newcommand{\beq}{\begin{equation}} 
\newcommand{\eeq}{\end{equation}} 
\newcommand{\beqa}{\begin{eqnarray}} 
\newcommand{\eeqa}{\end{eqnarray}}

\begin{document}

\voffset=-.01cm

\title{Aspects of Neutrino Production in Supernovae}
\author{Todd A.~Thompson\footnote{\uppercase{L}yman \uppercase{S}pitzer \uppercase{J}r. \uppercase{F}ellow}}
\address{
Department of Astrophysical Sciences\\
Princeton University\\ 
Peyton Hall --- Ivy Lane\\ 
Princeton, New Jersey, 08544 USA \\
E-mail: thomp@astro.princeton.edu
}

\maketitle

\abstracts{I discuss neutrino production in supernovae (SNe) and the
detection of both Galactic core collapse events and the
diffuse extra-galactic MeV neutrino background expected from 
the integrated history of star formation.  In particular, 
I consider what processes might affect our expectations for both.  
I focus on ``rapid'' rotation, defined as leading to millisecond initial neutron star 
spin periods.  Rotation affects the neutrino luminosity, the average
neutrino energy, the duration of the Kelvin-Helmholtz cooling epoch,
and the ratios of luminosities and average energies between neutrino 
species; it can strongly suppresses the $\bar{\nu}_e$ as well as 
$\nu_\mu$, $\bar{\nu}_\mu$, $\nu_\tau$, and
$\bar{\nu}_\tau$ fluxes relative to $\nu_e$.  As a result, depending on the prevalence of rapid
rotation in SN progenitors through cosmic time, this may affect 
predictions for the MeV neutrino background and the history of nucleosynthetic 
enrichment. I emphasize connections between the MeV neutrino background and 
tracers of the star formation rate density
at high redshift in other neutrino and photon wavebands.}

\section{Introduction}

When the iron core of a massive star collapses, the implosion is reversed at nuclear 
densities when nuclei dissociate into free nucleons.  The equation of state stiffens 
dramatically, driving a bounce shockwave into the supersonic infalling outer core.
The bounce shock stalls almost immediately as a result of neutrino losses, the ram 
pressure of the infalling material, and  the dissociation of nuclei into 
free nucleons across the shock.  After the shock stalls, a characteristic post-bounce 
accretion structure obtains that is quasi-steady-state.   The hot ($\sim$10\,MeV) 
newly-born ``proto''-neutron star (PNS) has a neutrinosphere radius of 
$\sim$80\,km.\footnote{Where the optical depth to $\nu_e$
neutrinos $\sim$2/3, an energy- and time-dependent quantity.}  Overlying the PNS is a subsonic 
accretion flow, bounded by the stalled stand-off shockwave at $\sim$$200$\,km.
Just above the PNS is a cooling layer and beyond that
a region of net neutrino heating (the ``gain'' region)
provided by both accretion luminosity as the matter falling through 
the bounce shock is incorporated into the PNS and by core neutrino luminosity as the PNS 
cools and deleptonizes.

The revival of the shock to an energy of order $10^{51}$\,ergs is the focus 
of supernova theory.  Any mechanism for this revival must fundamentally rely on 
the transfer of gravitational binding energy to the post-shock mantle.   

The ``neutrino mechanism''\cite{bethe}$^,$\cite{janka2001}
employs neutrino interactions  --- primarily 
$\nu_e n\leftrightarrow pe^-$ and $\bar{\nu}_e p\leftrightarrow n e^+$ ---
to transfer binding energy to the shocked matter in the post-bounce epoch. 
With the standard suite of microphysics and a solution
to the Boltzmann equation for all neutrino species, the neutrino mechanism 
fails in one spatial dimension; the stalled bounce shock remains trapped forever, 
accreting the overlying stellar progenitor.\cite{rampp2000}\cdash\cite{sumiyoshi2005}
Although some models show that multi-dimensional effects might be necessary for success 
of the neutrino mechanism,\cite{herant}\cdash\cite{fryer1999}
recent calculations employing more sophisticated neutrino 
transport fail to explode\cite{buras2003}
--- albeit marginally. In some cases, successes are 
obtained.\cite{buras2006}  However, the recent
calculation of Ref.~\refcite{burrows2006} suggests that energy deposition via neutrino 
interactions may be sub-dominant at late times with respect to acoustic heating 
generated by oscillations of the PNS generated by anisotropic 
accretion; the systematics of this new ``acoustic'' mechanism have yet to 
be elaborated.  Many recent complimentary works focus on 
the stability of the shockwave and the post-shock material,
and its importance for the mechanism.\cite{blondin}\cdash\cite{foglizzo} 

Quite apart from the details of the explosion mechanism, the total energy budget
dictating the character of neutron star birth is set by the gravitational binding 
energy of the neutron star:
\beq
E_{\rm bind}\approx3\times10^{53}M_{1.4}^2R_{10}^{-1}\,\,\,{\rm ergs},
\label{binding}
\eeq
where $M_{1.4}=M/1.4$\,M$_\odot$ and $R_{10}=R/10$\,km are the 
neutron star mass and radius, respectively.  
Theoretical models suggest\cite{burrows_lattimer}$^,$\cite{pons1999} and
detection of neutrinos from SN\,1987A confirm that a fraction of order
unity of this energy is radiated in $\sim10$\,MeV neutrinos on the Kelvin-Helmholtz 
timescale $\tau_{\rm KH}\sim10-100$\,s, long with respect to the collapse and 
explosion timescales.  Using these results, 
a number of studies have been made of the MeV neutrino background expected 
from SNe.\cite{ando2002}\cdash\cite{yuksel2005}
Indeed, the constraints on the background from SuperK\cite{malek2003} have become tight enough 
that, for example, a total energy radiated in $\bar{\nu}_e$ neutrinos of $E_{\rm bind}/3$
per SN with $\langle\varepsilon_{\bar{\nu}_e}\rangle>20$\,MeV is excluded.\cite{yuksel2005}
Estimates of the background must assume the total neutrino energy radiated per neutrino species per SN,
the average neutrino energy per species, and neutrino oscillation parameters. For light water detectors
like SuperK the dominant detection mode is single-species ($\bar{\nu}_e p\rightarrow n e^+$) and because
the cross section for this interaction is proportional to the neutrino energy squared 
the partitioning of energy, mixing, and the spectral shape are essential.  
The typical assumption is that the energy is partitioned equally between species,
that the spectrum is thermal,  and that the average energy is $\sim10-20$\,MeV.

Although the total energy budget (in all channels) per SN is dictated by equation (\ref{binding}),
it is interesting to consider what processes might affect the best estimates of the diffuse neutrino
background and the expected neutrino signal from the next Galactic SN at order unity. Potential processes 
are both microphysical and macrophysical/astrophysical.  
As an example of the former, it is possible 
that a change to the neutrino opacities or the equation of state for dense matter may alter the neutrino
luminosity  ($L_\nu$) and average energy ($\langle\varepsilon_\nu\rangle$)
during the cooling epoch, while the total energy radiated
per SN is unaltered.\cite{pons1999}   Alternatively, as an example of astrophysical uncertainty, one may include the 
assumption of a universal IMF with constant (neutron-star-producing) SN rate per solar mass 
per year of star formation as a function of redshift, environment, and metallicity.
In addition, there are a number of potential macrophysical effects that might modify our expectations
at first order, including rapid rotation of the massive progenitor's iron core just before collapse.\cite{heger2000}

Millisecond rotation of the PNS is interesting for several reasons. First,
it changes $L_\nu$ and $\langle\varepsilon_\nu\rangle$ and their ratios between species. 
This strongly effects the neutrino signature of core collapse and it may affect
nucleosynthesis in the inner SN ejecta by altering the ratio of the 
$\nu_e$ to $\bar{\nu}_e$ fluxes during explosion.\cite{pruet2005}$^,$\cite{frolich2005} 
Second, rapid rotation may be accompanied by 
more significant gravitational wave emission than non-rotating progenitors, thereby
opening up a new channel of emission for a fraction of $E_{\rm bind}$.  Third, 
rapid rotation may affect the morphology and nucleosynthesis of the remnant.\cite{tcq}$^,$\cite{bucciantini2006}
Finally, rapid rotation at birth has been theoretically motivated by the 
existence of magnetars.  Ref.~\refcite{duncan_thompson} argue that millisecond
rotation may be neceesary for production of the very high magnetic field
strengths ($\sim$$10^{15}$G) inferred for the magnetar neutron star subclass.
Although uncertain, simple estimates imply that $>10$\% of all SNe produce magnetars.\cite{woods_thompson}

In \S\ref{section:rotation}, I summarize results\cite{tqb} from a set of 1D simulations of 
core collapse, including rotation in a parameterized way, and I briefly
discuss the implications of these models for neutrino detection.  In \S\ref{section:budget}
I discuss the neutrino background and its connection to other observable backgrounds.

\section{Neutrinos from Rotating Core Collapse}
\label{section:rotation}

Details of the models presented here are given in Ref.~\refcite{tqb}.
A rotating 15\,M$_\odot$ progenitor (model ``E15A'') from Ref.~\refcite{heger2000}
and a set of 11\,M$_\odot$ progenitors from Ref.~\refcite{woosley_weaver} with imposed initial
rotation profiles are calculated.  The imposed profile is taken to be 
$\Omega(r)=(2\pi/P_0)[1+(r/R_\Omega)]^{-2}$, where $R_\Omega=1000$\,km; thus, the iron core is
in roughly solid-body rotation out to $R_\Omega$.
The 11\,M$_\odot$ models span spin periods of $1\lesssim P_0\lesssim 10$\,s ---
the former corresponding to near-, but sub-Keplerian rotation after collapse and 
the latter yielding results virtually indistinguishable from non-rotating models.
The $\Omega(r)$ profile for the 15\,M$_\odot$ model most resembles the $P_0=2$\,s
11\,M$_\odot$ progenitor (see Fig.~1 of Ref.~\refcite{tqb}).

\begin{figure}[t]
\centerline{\epsfxsize=4.5in\epsfbox{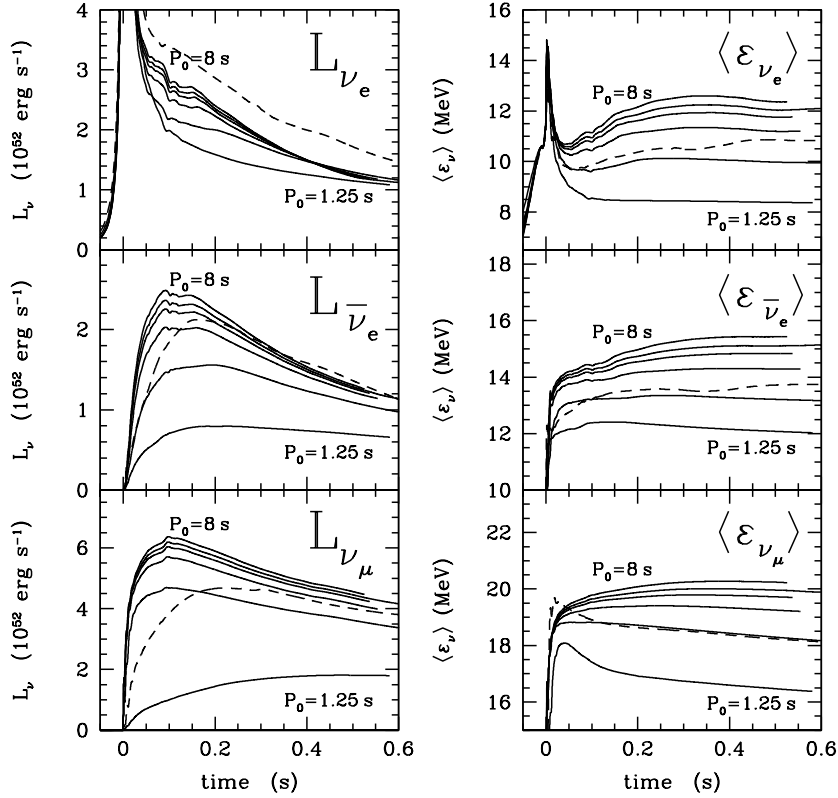}}   
\caption{$L_\nu$ ({\it left column})
and $\langle\varepsilon_\nu\rangle$ 
({\it right column}) as a function of time after bounce for the 11\,M$_\odot$
progenitor with initial core spin periods of $P_0=1.25$, 2, 3, 4, 5, and 8\,s
({\it solid lines}) and for the model E15A ({\it dashed lines}). 
200 ms after bounce the lowest $L_\nu$ and $\langle\varepsilon_\nu\rangle$ correspond
to the shortest initial spin period.  This hierarchy is preserved
at all times and in all panels except in the case of $L_{\nu_e}$
400\,ms after bounce and at breakout.   The $\nu_e$ breakout pulse 
is not shown in the upper-left panel. It is largest
for the fastest rotator, with peak $L_{\nu_e}$
$\sim$30\% larger than that for the slowest rotator ($\approx2.4\times10^{53}$\,ergs\,s$^{-1}$).
\label{plot:ltet}}
\end{figure}

\begin{figure}[t]
\centerline{\epsfxsize=4in\epsfbox{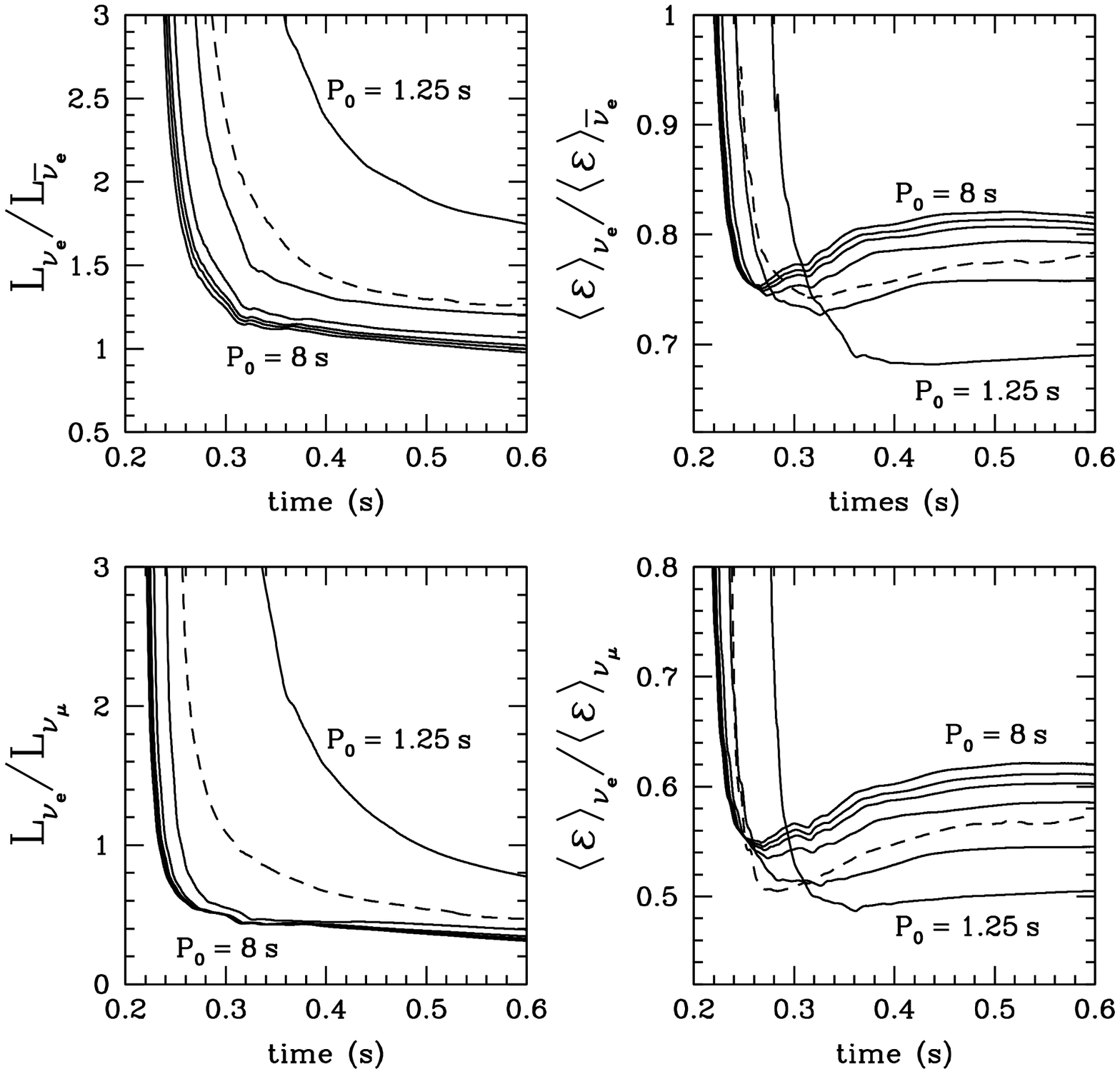}}   
\caption{Ratios of $L_\nu$s ({\it left column}) and 
$\langle\varepsilon_\nu\rangle$s ({\it right column}) from Fig.~\ref{plot:ltet}. 
\label{plot:llt}}
\end{figure}

Figure \ref{plot:ltet} shows $L_\nu$ and $\langle\varepsilon_\nu\rangle$ 
as a function of time after bounce for each of our rotating models.  
None of the models reach a centrifugal barrier and none explode.
Higher initial rotation rates yield lower core temperatures after bounce for the PNS.
On average, after $\nu_e$ breakout, this effect produces lower core $L_\nu$ and $\langle\varepsilon_\nu\rangle$ for shorter $P_0$.
The fractional differences in $\langle\varepsilon_\nu\rangle$ between the model with $P_0=1.25$ s and
a non-rotating model 200 ms after bounce are approximately 15\%, 17\%, and 30\% 
for $\nu_\mu$, $\bar{\nu}_e$, and $\nu_e$, respectively.  The same comparison for $L_\nu$ 
yields fractional differences of 75\%, 63\%, and 33\%
for $L_{\nu_\mu}$, $L_{\bar{\nu}_e}$, and $L_{\nu_e}$, respectively.
The difference in $L_{\nu_\mu}$ at 100\,ms after bounce between 
our slowest and fastest rotators is a factor of $\sim6$.  
Despite the fact that model E15A has rapid rotation, it has a much larger $L_{\nu_e}$ after breakout than, for
example, the $P_0=2$ s 11 M$_\odot$ model, even though they have similar initial $\Omega(r)$ profiles.  
This is due to the extended density profile of model E15A relative 
to the 11 M$_\odot$ model and the associated larger accretion luminosity after bounce.  

Figure \ref{plot:llt} shows the ratio of $L_{\nu_e}$ and $\langle\varepsilon_{\nu_e}\rangle$
to $L_\nu$ and $\langle\varepsilon_\nu\rangle$ for each species as a function of time in each 
rotating model. The total luminosity in
$\mu-$ and $\tau-$type neutrinos, $L_{\nu_\mu}$, is suppressed
by rapid rotation: for slow rotation
($P_0=8$\,s) the ratio $L_{\nu_e}/L_{\nu_\mu}\approx0.3$ at $0.5$s after
bounce, whereas for $P_0=1.25$\,s,  $L_{\nu_e}/L_{\nu_\mu}\approx0.8$.
A similar enhancement of $L_{\nu_e}$ relative to $L_{\bar{\nu}_e}$
is also seen in the upper left panel.
Although Figure \ref{plot:ltet} shows that $\langle\varepsilon_{\nu_e}\rangle$
is decreased on average by rotation, the right panels here show that 
both $\langle\varepsilon_{\bar{\nu}_e}\rangle$ and $\langle\varepsilon_{\nu_\mu}\rangle$
are decreased yet more.  The two upper panels may be particularly important 
for the nucleosynthesis of inner ejecta in the gain region, should an
explosion occur.  Several studies address how the electron fraction and
nucleosynthesis are determined by $L_{\nu_e}/L_{\bar{\nu}_e}$ and 
$\langle\varepsilon_{{\nu}_e}\rangle/\langle\varepsilon_{\bar{\nu}_e}\rangle$ 
as the explosion commences.\cite{pruet2005}$^,$\cite{frolich2005}  
A general study with these ratios motivated by Figure \ref{plot:llt} may constrain the 
frequency of SN events with rapid rotation.

\begin{figure}[t]
\centerline{\epsfxsize=3.6in\epsfbox{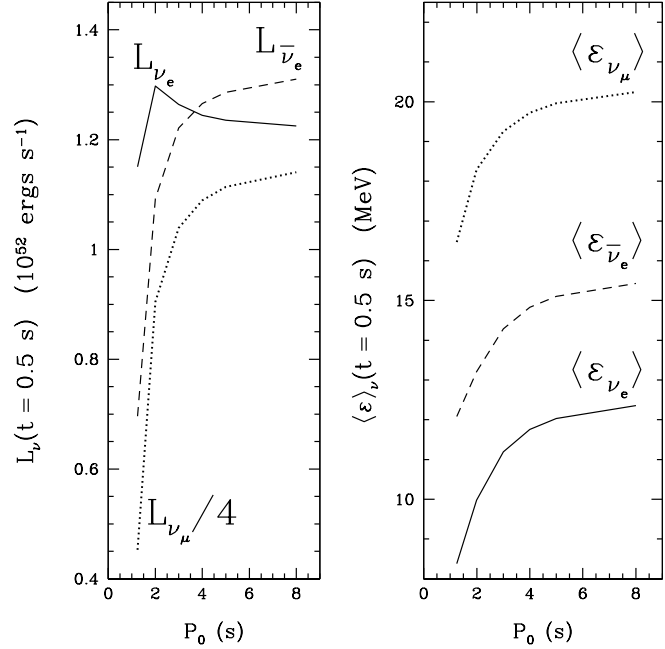}}   
\caption{$L_\nu$ 
and $\langle\varepsilon_\nu\rangle$,
at $t=0.5$\,s after bounce, as a function
of $P_0$ (see Figs.~\ref{plot:ltet} \& \ref{plot:llt}).
Although $L_{\nu_e}$ is fairly constant, $L_{\bar{\nu}_e}$ and  $L_{\nu_\mu}$ decrease sharply
with more rapid initial core rotation. 
\label{plot:o}}
\end{figure}

Figure \ref{plot:o} shows $L_\nu$ and $\langle\varepsilon_\nu\rangle$
at $0.5$\,s after bounce for the 11\,M$_\odot$ progenitor as a function of $P_0$.  
In the effectively non-rotating case $P_0=8$\,s,  
$L_{{\nu}_e}:L_{\bar{\nu}_e}:L_{{\nu}_\mu}/4::1:1.1:0.95$.  As Figure \ref{plot:o}
makes clear, this equality between neutrino species does not persist when
rotation is rapid; for $P_0=1.25$\,s, $L_{{\nu}_e}:L_{\bar{\nu}_e}:L_{{\nu}_\mu}/4::1:0.6:0.4$.
Although not explored systematically as a function of $P_0$, 
multi-D simulations of rotating collapse also exhibit suppression of $\bar{\nu}_e$ and $\nu_\mu$ at 
the level described here.\cite{fryer_heger}$^,$\cite{dessart2006}

\begin{figure}[ht]
\centerline{\epsfxsize=3.4in\epsfbox{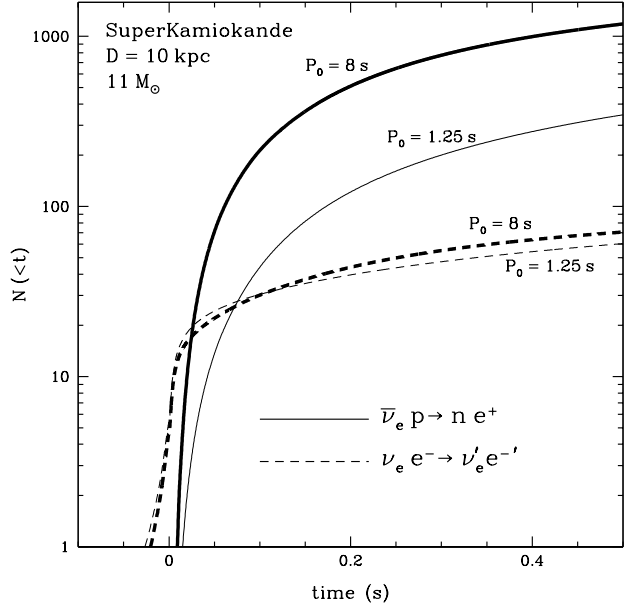}}   
\caption{Integrated number of events in SuperK $N(<t)$
for a SN at 10\,kpc as a function of time
after bounce.  Only the $P_0=8$s ({\it heavy lines}) and $P_0=1.25$s ({\it thin lines}) models 
are shown.  Event rates for the reactions
$\nu_e e^-\rightarrow\nu_e^{\prime} e^{-\,\prime}$ ({\it dashed lines}) and
$\bar{\nu}_e p\rightarrow p e^+$ ({\it solid lines}) are shown separately.
In the rapidly rotating model the $\bar{\nu}_e$ signal is strongly suppressed  
and the prompt $\nu_e$ signal persists for $\approx0.05$s before being swamped by the 
$\bar{\nu}_e$s, much longer than for the $P_0=8$s model. 
\label{plot:d}}
\end{figure}

The strong suppression of both the $\bar{\nu}_e$ and $\nu_\mu$ luminosity and average energy 
is important for the detection of neutrinos from the next Galactic SN.
Figure \ref{plot:d} shows the integrated number of events expected in SuperK
from the 11\,M$_\odot$ progenitors with $P_0=8$\,s and $1.25$\,s at a distance of 10\,kpc
from the interactions $\bar{\nu}_e p\rightarrow n e^+$ and $\nu_e e^-\rightarrow\nu_e^\prime e^{-\,\prime}$.
Neutrino oscillations within the SN progenitor envelope, while the neutrinos
are in transit, or within the Earth are neglected.  The number rate of neutrinos 
detected is $\dot{N}\propto [L_\nu/\langle\varepsilon_\nu\rangle]\sigma$, where 
$\sigma$ is the cross section.  Because $\sigma\propto\langle\varepsilon_{\nu_e}\rangle$ 
for inelastic $\nu_e-e^-$ scattering and because $L_{\nu_e}$ is roughly
independent of $P_0$ (Fig.~\ref{plot:o}), the total number of
$\nu_e - e^-$ scattering events in SuperK is approximately independent of $P_0$,
although the $\nu_e$ breakout burst is somewhat larger in the rapidly rotating case, $P_0=1.25$\,s.
However, the dominant charged-current interaction  $\bar{\nu}_e p\rightarrow n e^+$
has $\sigma\propto\langle\varepsilon_{\bar{\nu}_e}^2\rangle$ so that for this
process $\dot{N}\propto L_{\bar{\nu}_e}\langle\varepsilon_{\bar{\nu}_e}\rangle$.
Thus, the strong decrease in both $L_{\bar{\nu}_e}$ and 
$\langle\varepsilon_{\bar{\nu}_e}\rangle$ with $P_0$ (Fig.~\ref{plot:o}) 
leads to a very large decrease in $N(<t)$: 0.5\,s after bounce the
ratio between the total number of neutrino events for the two models is $\approx1190/350\approx3.4$.

\section{The Energy Budget of the Universe}
\label{section:budget}

A simple estimate for the 
contribution of SNe to the MeV neutrino background can be made by 
relating the total energy expected in neutrinos from each 
SN to the star formation rate, $\dot{M}_\star$,
which is related to the  total IR luminosity ($L_{\rm TIR}$ [$8-1000$]$\mu$m)
by $L_{\rm TIR}=\epsilon\dot{M}_\star c^2$, where $\epsilon$
is an IMF-dependent constant.  Assuming that the 
SN rate per unit star formation $\Gamma_{\rm SN}$
is a constant fraction of $\dot{M}_\star$ and that the
total energy radiated is $E_\nu^{\rm tot}=10^{53.5}$\,ergs per SN,
one finds that (averaging over a suitably large number of systems
or time)
\beq
L_\nu\approx3L_{\rm TIR}E_{53.5}\beta_{17},
\label{nu_ir}
\eeq
where $L_\nu$ is the total neutrino luminosity and 
$\beta_{17}=(\Gamma_{\rm SN}/\epsilon)/17$\,M$_\odot$
has only a weak dependence on the assumed IMF because massive stars
dominate photon and neutrino production.
The total specific intensity in MeV SN neutrinos expected from the
star formation history of the universe can then be scaled from 
the integral over the comoving star formation rate density as a function of 
redshift, $\dot{\rho}_\star(z)$.
Taking $\Omega_m=0.3$ and $\Omega_\Lambda=0.7$ and using 
``RSF2'' for $\dot{\rho}_\star(z)$ from Ref.~\refcite{porciani_madau}, one finds that 
$\zeta = \int dz [\dot{\rho}_\star(z)/\dot{\rho}_\star(z=0)]/\{(1+z)^2[\Omega_m(1+z)^3+\Omega_\Lambda]^{1/2}\}\approx2$.
Thus, 
\beq
F_\nu^{\rm tot}\sim4\,(\zeta/2)\,\,\,\,\,{\rm 10\,MeV\,\,cm^{-2}\,\,s^{-1}\,\,sr^{-1}},
\label{background}
\eeq
where $H_0=71$\,km\,s$^{-1}$\,Mpc$^{-1}$ has been assumed 
and the dependences on $E_\nu^{\rm tot}$ and $\beta$ have been dropped.  The dominant
reaction for detection of relic neutrinos in SuperK is $\bar{\nu}_e p\rightarrow n e^+$. 
Assuming that $F_{\bar{\nu}_e}\approx F_\nu^{\rm tot}/6$, 
$\langle\varepsilon_{\bar{\nu}_e}\rangle\approx15$\,MeV, and accounting for the fact that 
SuperK's sensitivity to the diffuse background is maximized for 
$\varepsilon_{\bar{\nu}_e}\sim20$\,MeV, considerably larger than $\sim$$15/(1+z)$\,MeV for $z\approx1$, 
one finds an event rate in SuperK of $\sim$$1$ yr$^{-1}$, in accord with more
complete estimates.\cite{ando2002}\cdash\cite{yuksel2005}   
As Figures \ref{plot:o} \& \ref{plot:d} make clear, if a large
fraction of all SNe are born rapidly rotating, then the background estimate
is significantly decreased because of the strong decrease in 
$\langle\varepsilon_{\bar{\nu}_e}\rangle$ and $\langle\varepsilon_{\bar{\nu}_\mu}\rangle$ with decreasing $P_0$.

As implied by equations (\ref{nu_ir}) and (\ref{background}), the massive stars that generate
MeV neutrinos also produce a  corresponding total IR background of
$F_{\rm TIR}^{\rm tot}\approx2\times10^{-5}\,\,{\rm ergs\,\,cm^{-2}\,\,s^{-1}\,\,sr^{-1}}
\approx20\,\,{\rm nW\,\,m^{-2}\,\,sr^{-1}}$.  The SNe that accompany this star formation
also accelerate cosmic ray electrons and protons to very high energies.  Inelastic collisions between cosmic ray protons
and gas in the ISM of star-forming galaxies produce $\pi^{+,-}$ and $\pi^0$, which subsequently decay
to $e^{-,+}$ and high-energy neutrinos, and $\gamma$-rays, respectively.  The
primary electrons and secondary electrons/positrons  suffer
synchrotron losses in the host galactic magnetic field.  Inverse Compton and bremsstrahlung
losses likely also contribute significantly to cooling.\cite{thompson2006_mag}  
The observed tight linear correlation
between the IR and radio luminosity of star-forming and starburst galaxies implies that 
the contribution to the IR background from star formation comes together with radio emission
at the level $\nu I_\nu({\rm radio})\approx3\times10^{-11}\,\,\,{\rm ergs\,\,cm^{-2}\,\,s^{-1}\,\,sr^{-1}}$,
assuming a flat cosmic ray spectrum --- consistent with $\sim$$5$\% of the $10^{51}$\,ergs
of asymptotic kinetic energy of each SN going into cosmic rays.\cite{thompson2006_back}
Recent work suggests that inelastic $p-p$ collisions in the dense ISM of 
starburst galaxies may contribute significantly to the diffuse GeV $\gamma$-ray and 
GeV$-$TeV $\nu_\mu$-neutrino background at the level of
$\sim$$10^{-7}\,\,\,{\rm GeV\,\,cm^{-2}\,\,s^{-1}\,\,sr^{-1}}$.\cite{thompson2006_back}$^,$\cite{loeb_waxman}
Thus, stars that produce SNe may dominate --- or contributor importantly to --- 
the neutrino (MeV \& GeV$-$TeV), $\gamma$-ray, IR, and radio backgrounds.

\section*{Acknowledgments}

I thank Adam Burrows, Eliot Quataert, and Eli Waxman for collaboration and many 
stimulating conversations.

\end{document}